\documentclass[twocolumn,superscriptaddress,showpacs,floatfix,amsmath,amssymb]{revtex4-1}

\usepackage{xcolor}
\usepackage{graphicx}
\usepackage[dvipdfmx,colorlinks=true,linkcolor=blue,citecolor=blue,urlcolor=blue]{hyperref}
\usepackage[utf8]{inputenc}

\newcommand{\bYb}{${}^{174}$Yb}
\newcommand{\bEr}{${}^{168}$Er}

\newcommand{\cm}{\mathrm{cm}}

\newcommand{\nm}{\mathrm{nm}}
\newcommand{\ms}{\mathrm{ms}}
\newcommand{\s}{\mathrm{s}}
\newcommand{\G}{\mathrm{G}}

\newcommand{\Hz}{\mathrm{Hz}}

\newcommand{\uK}{\mu\mathrm{K}}
\newcommand{\nK}{\mathrm{nK}}
\newcommand{\W}{\mathrm{W}}
\newcommand{\PSD}{\mathrm{PSD}}
\newcommand{\lossunit}{\cm^6\,\s^{-1}}

\begin{document}

\title{Realization of a quantum degenerate mixture of highly magnetic and nonmagnetic atoms}

\author{F.~Sch\"{a}fer}
\email{schaefer@scphys.kyoto-u.ac.jp}
\affiliation{Department of Physics, Graduate School of Science, Kyoto University, Kyoto 606-8502, Japan}

\author{Y.~Haruna}
\affiliation{Department of Physics, Graduate School of Science, Kyoto University, Kyoto 606-8502, Japan}

\author{Y.~Takahashi}
\email{takahashi.yoshiro.7v@kyoto-u.ac.jp}
\affiliation{Department of Physics, Graduate School of Science, Kyoto University, Kyoto 606-8502, Japan}

\date{\today}

\begin{abstract}
	We report on the experimental realization of a bosonic quantum degenerate
	mixture of highly magnetic \bEr\ and nonmagnetic \bYb. Quantum degeneracy is
	reached by forced evaporation in an all-optical trap. Formation of the two
	Bose-Einstein condensates is confirmed by analysis of the cloud shape and
	the observed inversions of the aspect ratios. The results open a path for
	possible new experiments on magnetic and nonmagnetic impurity physics as
	well as on the quantum chaotic behavior of Feshbach resonances and their
	dependencies on minor variations of the reduced masses.
\end{abstract}

\maketitle

\section{Introduction}
\label{sec:intro}

The research field of ultracold atoms has become a vast and rich one. The
number of atomic species that have been brought to quantum degeneracy spans
across the entire periodic table of elements, reaching from the lightest
ones~\cite{fried_bose-einstein_1998} to very heavy
species~\cite{takasu_spin-singlet_2003}, and is by now too large to list them
all individually here. In addition to single-species experiments, two-element
experiments have also gained considerable attention. Starting with an
ultracold Cs-Li mixture~\cite{mosk_mixture_2001}, this field also evolved in
many directions, again covering a range from large mass-imbalance Yb-Li
mixtures~\cite{hansen_quantum_2011} to highly magnetic and anisotropic Er-Dy
mixtures~\cite{trautmann_dipolar_2018}. This diversity is not redundancy but a
richness of possibilities, each one looking at the fundamental questions
stemming from quantum physics, many-body physics, quantum simulation and
quantum computation at slightly angles.

It is in this spirit that we here introduce another member to the family of
experimentally realized quantum degenerate mixtures: a mixture of bosonic
erbium (\bEr) and bosonic ytterbium (\bYb). Without going to the extremes of
either large mass imbalances or nearly overwhelmingly complicated interspecies
interaction potentials, this mixture might be ideal for exploring physical
phenomena in-depth and systematically with reasonable computational efforts,
as first proposed and theoretically described by Kosicki, Borkowski, and
{\.Z}uchowski~\cite{kosicki_quantum_2020}. In short, while quantum chaotic
behavior could be confirmed experimentally in, e.g., the complex dynamics of
highly magnetic and highly anisotropic Er-Er
collisions~\cite{frisch_quantum_2014}, it is believed that evidence of quantum
chaos may also be found in the significantly simpler Er-Yb collisional system.
In that case, one of the collisional partners, Yb, is a
comparatively-easy-to-model spin-singlet atom, and the interspecies
interactions can be derived from \emph{ab initio}
calculations~\cite{kosicki_quantum_2020}. This greatly helps in identifying
the driving mechanisms, such as anisotropic interactions, that contribute to
the chaotic behavior. Furthermore, the two species have similarly large masses
while also offering various isotopes each for use in the experiments, Er-Yb
mixtures offer a unique possibility to study the effects of minute
reduced-mass differences on the Feshbach spectra. The expected shifts of the
interspecies Feshbach resonance positions should be well controlled and might
be sensitive to temporal changes in the proton-to-electron mass
ratio~\cite{ketterle_evaporative_1996}.

In addition, this quantum degenerate mixture is unique in that one of its
constituents is a highly magnetic atom, \bEr, with a magnetic moment of $7\,
\mu_B$ whereas the other one, \bYb, is nonmagnetic. This offers the
possibility of studying the magnetic-nonmagnetic impurity problem at vanishing
interspecies interactions by means of a Feshbach resonance, especially when
loaded into an optical lattice. There, the additional existence of narrow and
ultranarrow optical transitions for both
Er~\cite{patscheider_observation_2021} and
Yb~\cite{yamaguchi_high-resolution_2010} would provide a powerful tool for
such investigations.

In the following we will first describe the details of our experiment relevant
to the formation of the quantum degenerate \bEr-\bYb\ mixture
(Sec.~\ref{sec:experiment}). We then present the experimental data and our
evidence for achieving simultaneous Bose-Einstein condensates (BECs;
Sec.~\ref{sec:results}) and conclude with a short discussion of these results
(Sec.~\ref{sec:discussion}).

\section{Experiment}
\label{sec:experiment}

\begin{figure}[tb]
	\centering
	\includegraphics[width=8cm]{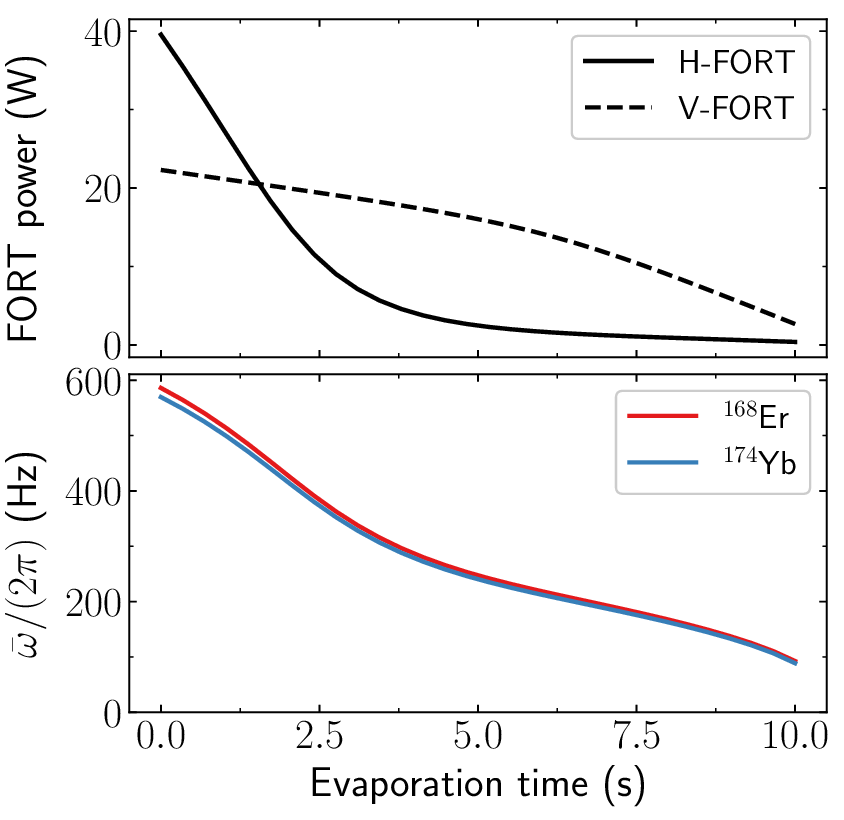}
	\caption{Trap laser powers and trap frequencies during the forced
		evaporation process. Top: The evaporation in the two crossed FORT lasers
		requires $10\,\s$. During that time the power of the H-FORT (solid line)
		is gradually decreased from roughly $40$ to $0.4\,\W$ at the atoms. The
		V-FORT (dashed line) is for better lateral confinement initially kept
		closer to the initial power of about $22\,\W$ and then reduced to about
		$2.7\,\W$. Bottom: The calculated geometric mean trap frequencies for Er
		(upper, red line) and Yb (lower, blue line). Due to the similar masses and
		polarizabilities the trap frequencies are quite similar. The trap
		frequencies show a nearly linear reduction in time, a surprising result of
		the independent evaporation ramp optimization that maximized the Yb BEC
		production efficiency.
	}
	\label{fig:fig1}
\end{figure}

The setup of the experiment and the experimental sequence are largely as in
our previous works~\cite{schafer_feshbach_2022}, and we will here only
highlight the key aspects and changes relevant to the present work. After
first loading the \bYb\ magneto-optical trap (MOT) to saturation in $45\,\s$ a
rather small amount of \bEr\ is added with just $2\,\s$ of loading. Due to the
narrow-linewidth nature of the Er MOT the obtained Er sample is automatically
spin polarized into the lowest energy $m_J = -6$
state~\cite{frisch_quantum_2014, schafer_feshbach_2022}. Both species are then
transferred into an optical far-off resonant trap in the horizontal direction
(H-FORT) at $1064\,\nm$ where we typically obtain about $3 \times 10^5$ Er
atoms and $4 \times 10^6$ Yb atoms, both at an estimated temperature of
$80\,\uK$. This one-order-of-magnitude difference in atom numbers is to
facilitate the sympathetic cooling of Er by Yb which this experiment relies
on. That is, by forced evaporation of Yb the Yb sample has to not only cool
itself down but also remove the energy from the Er sample which is in good
thermal contact with Yb by means of frequent elastic collisions. Before
evaporation a second FORT beam in the vertical direction (V-FORT) at
$1070\,\nm$ is added for increased atom densities and thus evaporation
efficiencies in the resulting crossed FORT configuration. In the following
forced evaporation step, both FORT powers are gradually reduced over a period
of $10\,\s$ as shown in the top panel of Fig.~\ref{fig:fig1}.

The evaporation ramp shape of each FORT is a spline curve with, in addition to
the start and end points, two additional in-between control points. The ramp
shapes and their duration have been optimized for optimal evaporation of \bYb.
During the first $4\, \s$ of evaporation a magnetic bias field of $1.6\, \G$
is applied in the vertical direction that is then reduced to $0.4\, \G$ for
the remainder of the evaporation ramp. This choice is to ensure that the
initial spin polarization of \bEr\ is maintained throughout the evaporation
while also having optimal interspecies collisional properties and low Er-Er
inelastic collisional losses~\cite{frisch_quantum_2014}.

Due to both the similar masses of \bEr\ and \bYb\ and the similar
polarizabilities of the two species at the wavelengths of the FORT lasers the
resulting traps are quite similar. As expressed in the bottom panel of
Fig.~\ref{fig:fig1} the geometric mean trap frequencies start at about $2\pi
\times 600\,\Hz$ and then decrease in a nearly linear fashion to about $2\pi
\times 100\,\Hz$ after $10\,\s$. The final trap frequencies are about
$(\omega_x, \omega_y, \omega_z) = 2\pi \times (55, 70, 190)\,\Hz$ where
$\omega_z$ is in the vertical direction. From earlier works it is known that
the \bYb-\bYb\ scattering length is $105\,a_0$, where $a_0$ is the Bohr
radius~\cite{kitagawa_two-color_2008}. The \bEr-\bEr\ scattering length was
measured to be between $150$ and $200\,a_0$ where, however, additional effects
due to dipole-dipole interactions should also be considered
carefully~\cite{aikawa_bose-einstein_2012}. The \bEr-\bYb\ interspecies
scattering length is as of yet unknown.

\section{Results}
\label{sec:results}

\begin{figure}[tb!]
	\centering
	\includegraphics[width=8cm]{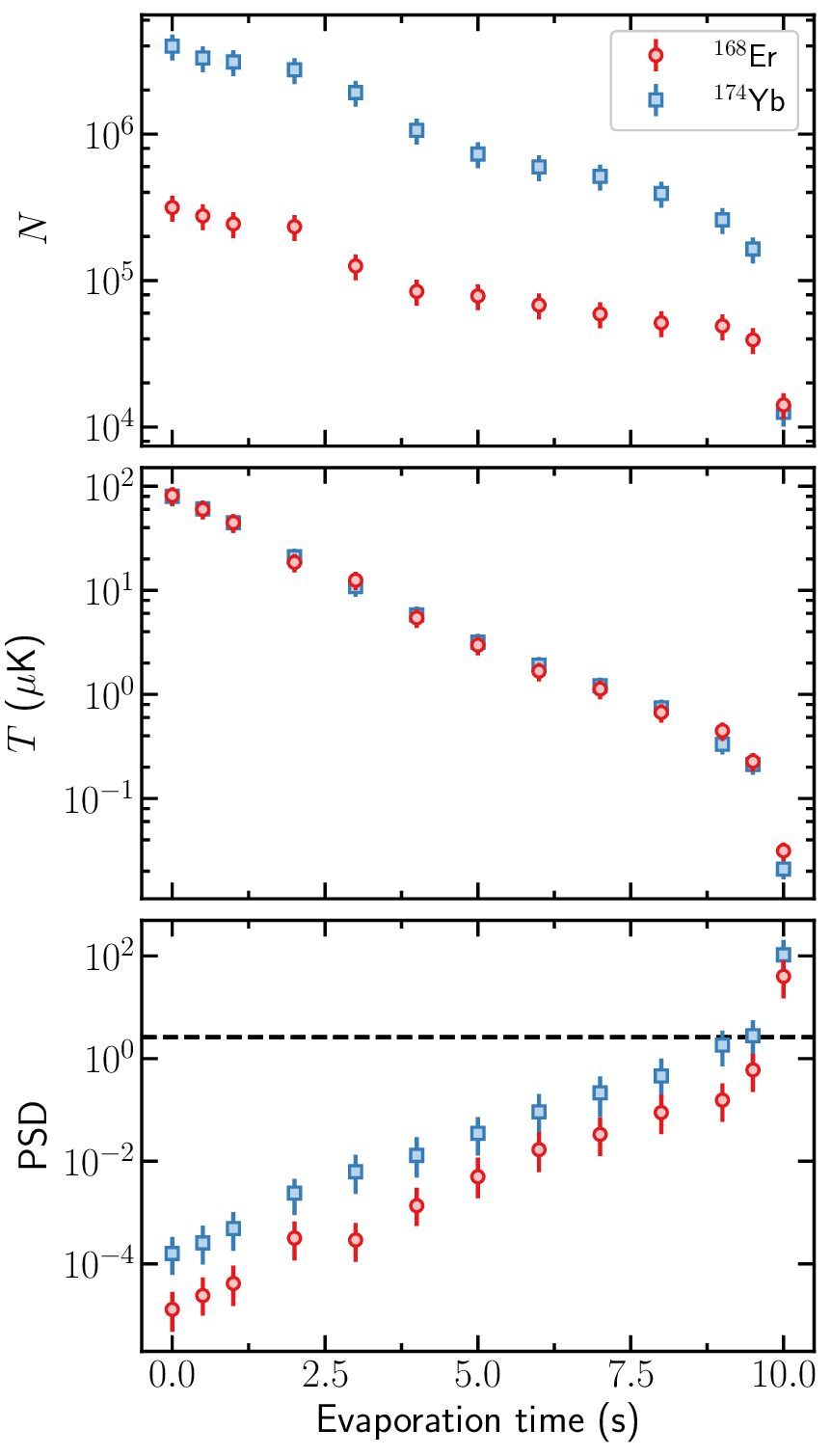}
	\caption{Development of atom numbers (top), temperatures (middle) and
		phase-space densities (bottom panel) during evaporation of the \bEr-\bYb\
		mixture. Initially, about one order of magnitude more Yb is loaded into
		the crossed FORT as it serves as the main coolant of the mixture. The
		close agreement of the Er and Yb temperatures demonstrates a very
		efficient interspecies thermalization dynamics, a necessary prerequisite
		for efficient sympathetic cooling of Er by Yb. Double quantum degeneracy
		of the sample is reached close to the end of the evaporation ramp where
		the phase-space densities rise above the critical threshold value of about
		$2.6$ (black dashed line in bottom panel). The data points are averages
		over typically six independent renditions of the experiment, and the error
		bars take the uncertainties in the atom number and temperature
		determinations as well as the estimated possible errors in the assumed
		trap frequencies into account. See the main text for additional comments
		on the data at $10\,\s$.
	}
	\label{fig:fig2}
\end{figure}

We now address the development of the atom numbers $N$, the temperatures $T$,
and the phase-space density (PSD) during the evaporation process, which is
shown in Fig.~\ref{fig:fig2}. As expected, at the expense of mostly Yb atoms
the overall temperature of the mixture decreases. It is important to note the
high fidelity with which the temperature of Er follows the temperature of Yb,
highlighting the very good thermal contact between the two species and
allowing for successful sympathetic cooling of Er. Correspondingly, the PSD
increases by more than four orders of magnitude. After about $9.5\,\s$ of
evaporation it reaches the critical value of about $2.6$, the onset of quantum
degeneracy into a BEC. This is further evidenced by the apparent discontinuity
of the observed quantities at $10\,\s$. This change is most probably due to
the phase transition from a thermal cloud to a BEC where a Gaussian fit is no
longer the appropriate description of the actual density distribution but
instead highlights the sudden ``compactification'' of the sample.

\begin{figure}[tb!]
	\centering
	\includegraphics[width=8cm]{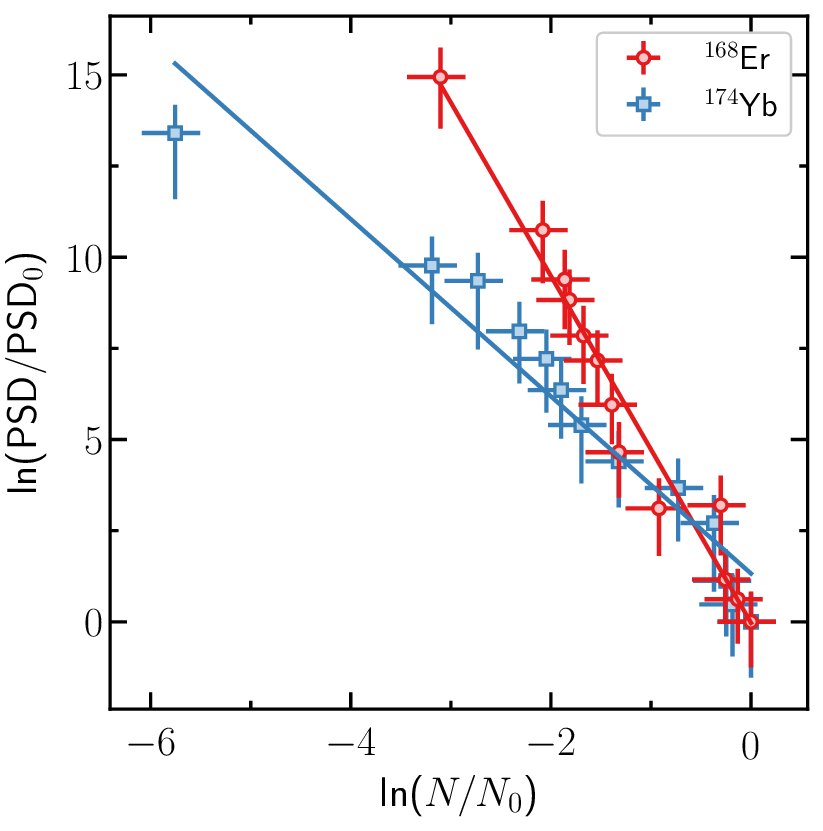}
	\caption{Development of the normalized phase-space densities ($\PSD/\PSD_0$)
		and the normalized atom numbers ($N/N_0$) during evaporation. $\PSD_0$ and
		$N_0$ are the respective values at the beginning of the evaporation ramp.
		The error bars account for uncertainties in the cloud and trap parameters.
		The trajectories for \bEr (red circles) and \bYb (blue squares) can be
		approximately described and fitted by power laws (straight lines) with
		exponents of $-4.8(3)$ for Er and $-2.4(2)$ for Yb. The apparent twofold
		more efficient evaporation of Er is clearly due to the efficient
		sympathetic cooling by Yb.
	}
	\label{fig:fig3}
\end{figure}

As a measure of the evaporation efficiency we look at the relationship of the
$\PSD$, normalized to its initial value $\PSD_0$, to the atom numbers $N$,
normalized to their initial values $N_0$. There, usually a power-law-like
behavior is observed, and $\gamma = -d\ln(\PSD/\PSD_0)/d\ln(N/N_0)$ serves as
a good indicator of evaporation efficiency~\cite{ketterle_evaporative_1996}.
In the case at hand (see Fig.~\ref{fig:fig3}), one obtains $\gamma = 4.8(3)$
for Er and $2.4(2)$ for Yb. Comparing this, e.g., with our earlier results in
the evaporation of an Yb-Li mixture~\cite{schafer_experimental_2018}, we find
a similar efficiency for Yb to that found before ($\gamma_{\rm Yb} = 2.9$ for
Yb-Li) but a reduced evaporation efficiency of the sympathetically cooled
species Er as compared with Li ($\gamma_{\rm Li} = 6.5$ in Yb-Li). This might
in part be due to the reduced effect of gravity on the very light Li resulting
in a significantly deeper trap and reduced losses for hotter Li atoms. The
sympathetic cooling of Er is not perfect in the sense that a reduction in the
Er atom number is also observed. It is rather that in this scenario of similar
trap depths for both species and also due to favorable Er-Er collisional
properties~\cite{aikawa_bose-einstein_2012}, Yb efficiently supports the
cooling process of Er. This is further evidenced by the fact that in our
experiment no Er BEC can be achieved by single-species evaporation due to an
insufficient number of Er atoms initially loaded into the FORT, while a
single-species-evaporated Yb BEC is possible.

\begin{figure}[tb!]
	\centering
	\includegraphics[width=8cm]{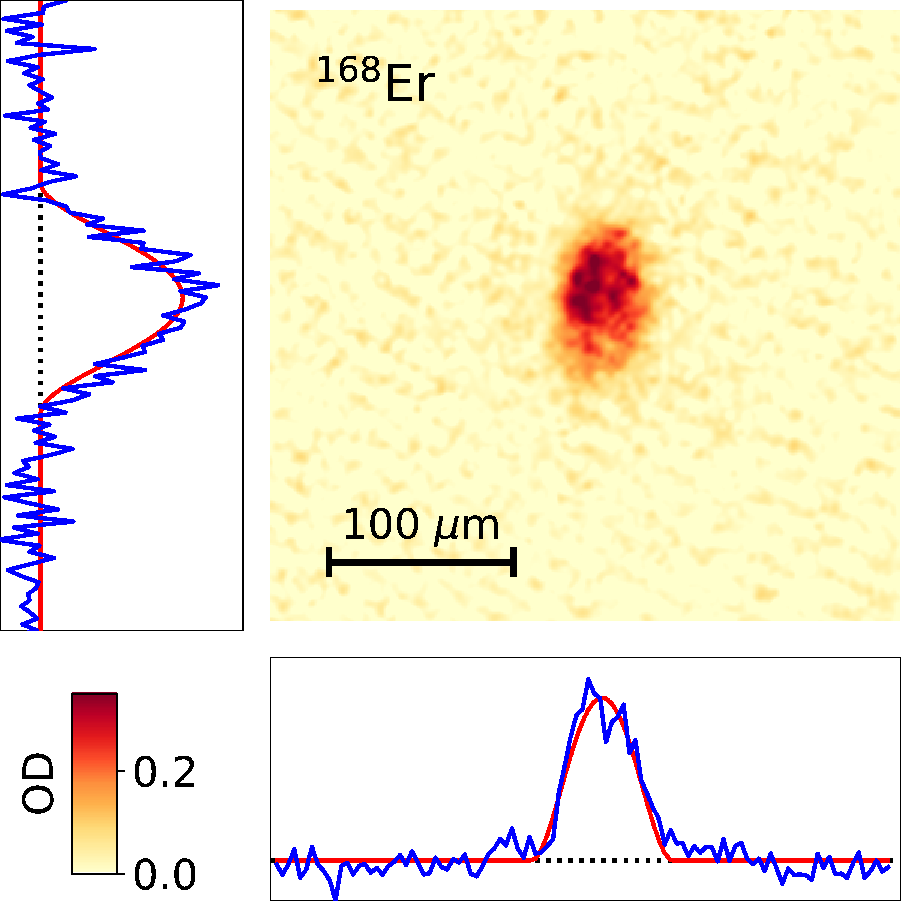}\\
	\vspace*{1mm}
	\includegraphics[width=8cm]{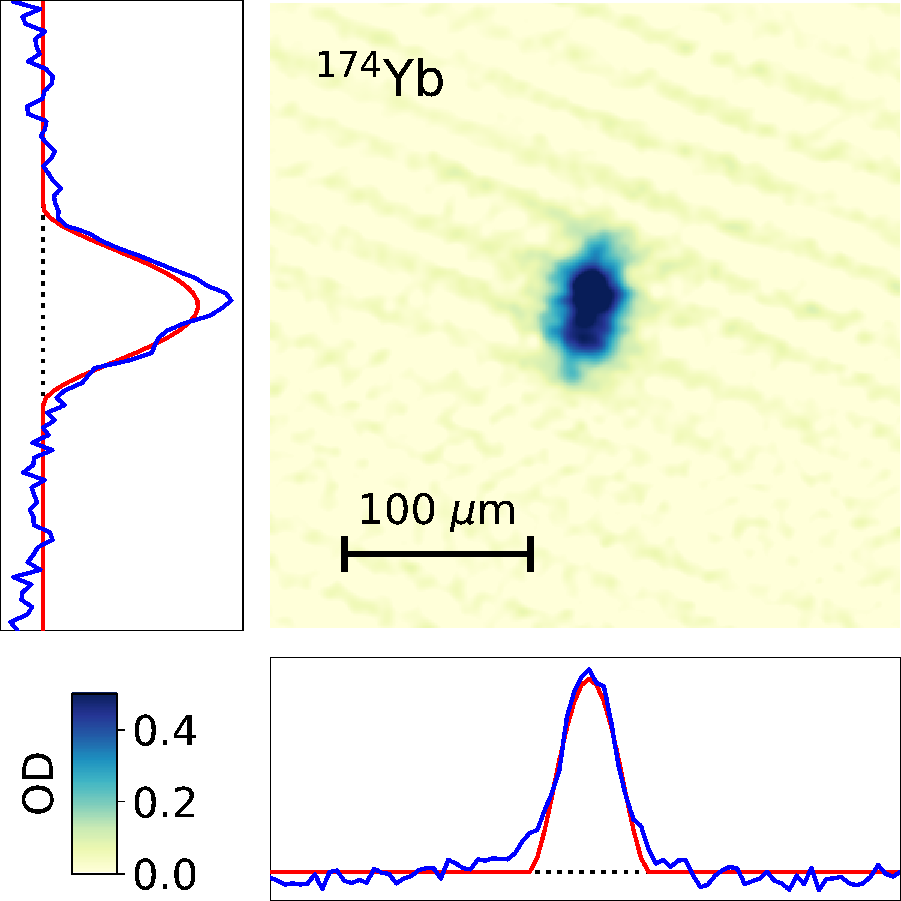}
	\caption{Absorption images of \bEr (top panel) and \bYb (bottom panel)
		Bose-Einstein condensates. Both images are the averages of 12 independent
		realizations of the experiment with absorption imaging after $22\ \ms$ of
		free expansion for both species. The colors indicate the recorded optical
		density (OD; see color bars). The additional graphs are the integrated
		column densities (ordinates are in a linear arbitrary units scale) of the
		data (blue) together with the fit results of Thomas-Fermi distributions
		(red). Even though the samples appear to be not fully quantum degenerate,
		the remaining thermal fractions are too low for reliable fitting with
		bimodal Thomas-Fermi--Gaussian distributions. The available fits indicate
		that there are about $1.3 \times 10^4$ Er atoms and $1.5 \times 10^4$ Yb
		atoms present in the final mixture.
	}
	\label{fig:fig4}
\end{figure}

We probe the dual-species BEC after $10\,\s$ of evaporation by standard
absorption imaging. Before releasing the sample from the optical trap we keep
the FORT lasers at their final values for an additional $500\,\ms$ to allow
for final thermalization of the two atom clouds. The trapping lasers are then
suddenly switched off, and the atoms are allowed to expand freely. After
$22\,\ms$ of expansion time both atomic species are imaged simultaneously in
absorption imaging on the strong transitions at $401\,\nm$ for Er and
$399\,\nm$ for Yb. The results are shown in Fig.~\ref{fig:fig4}. Both clouds
are reasonably well described by Thomas-Fermi distributions, indicative of
only small contributions from still thermal atoms. Indeed, tentative fits with
bimodal distributions (i.e., distributions including Gaussian-shaped thermal
contributions) seemed unrealistic, a situation typically experienced in the
case of only small thermal contributions. We find no appreciable
change in the cloud shapes in the case where one species is removed before
releasing the remaining atoms from the trap. This indicates that the Er-Yb
interspecies interactions are below the threshold beyond which immiscibility
effects are expected~\cite{pethick_bose-einstein_2008,
burchianti_dual-species_2020} and gives us the possibility to estimate an
upper bound~\cite{riboli_topology_2002} for the Er-Yb scattering length,
$a_{\rm ErYb}$: Assuming an \bEr\ scattering length of $a_{\rm Er} \le
200\,a_0$ and taking the known $a_{\rm Yb} = 105\,a_0$ it follows that
$|a_{\rm ErYb}| < 145\,a_0$.

\begin{figure}[tb]
	\centering
	\includegraphics[width=8cm]{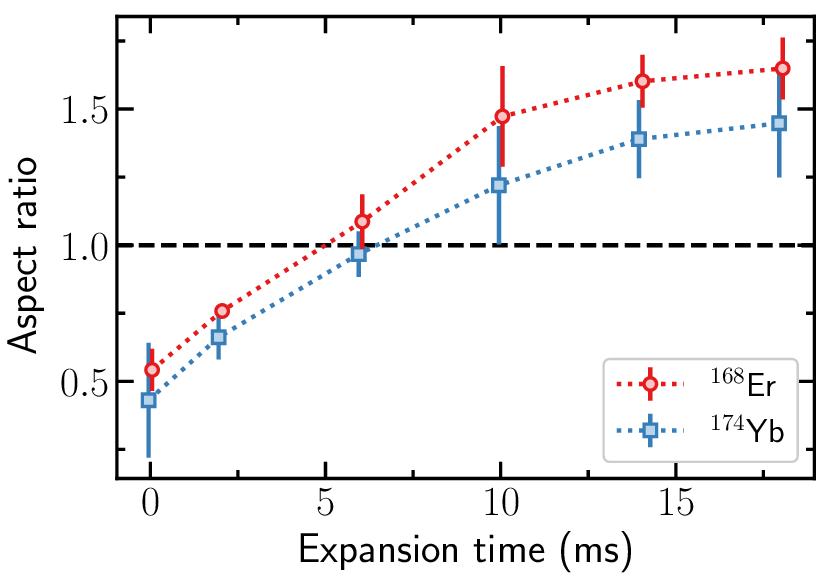}
	\caption{Development of the cloud aspect ratios for Er (red circles) and Yb
		(blue squares) during different expansion times. The aspect ratio is here
		defined as the ratio of cloud height over cloud width. It follows that at
		short expansions times the clouds still maintain their in-trap shape.
		After $5$--$6\ \ms$ the anisotropic expansion due to the stored
		interaction energy causes both clouds to become elongated in the vertical
		direction. This inversion of the aspect ratio is a hallmark feature of
		BECs that cannot be observed with thermal clouds. The dashed lines are
		intended as a guide to the eye.
	}
	\label{fig:fig5}
\end{figure}

The overall shape of the atom clouds in Fig.~\ref{fig:fig4} is not spherically
symmetric as would be expected of thermal samples. Instead they are elongated
in the vertical direction. A more systematic investigation of the dependence
of these aspect ratios on the expansion time is presented in
Fig.~\ref{fig:fig5}, where expansion times in the range $0$--$18\,\ms$ are
investigated. For both species an inversion of the aspect ratio at about
$5$--$6\,\ms$ of expansion time is observed. This agrees reasonably well with
a theoretically expected inversion point at about $4\, \ms$ based on the final
trapping frequencies~\cite{castin_bose-einstein_1996} and further corroborates
the quantum degenerate nature of the ultracold mixture of \bEr\ and \bYb.

Evaluating the stability of the obtained BEC mixture at $0.4\,\G$ we observe
lifetimes of the Er BEC in excess of $10\,\s$ while the lifetime of our Yb BEC
is generally lower, at about $2\,\s$, which is purely due to technical
limitations of our apparatus and not due to the presence of Er. Therefore, for
a more qualitative evaluation of the inelastic collisional processes, we
investigate the atom loss with a cold but still thermal mixture at about
$400\,\nK$. There, we find single-species three-body collision
rates~\cite{ye_observation_2022} for \bEr\ and \bYb\ of $3(1) \times 10^{-28}$
and $1.0(5) \times 10^{-28}\,\lossunit$, respectively. The additionally
observed interspecies three-body collisional losses are at $1.0(5)\times
10^{-28}\,\lossunit$ comparable to the intraspecies losses and well described
by Er-Yb-Yb interactions while Er-Er-Yb collisions are not compatible with our
observations. These low losses underline the good suitability of ultracold
\bEr-\bYb\ mixtures for future experiments.

\section{Discussion}
\label{sec:discussion}

First and foremost the work presented here confirms that a stable formation of
a \bEr-\bYb\ double BEC is not only theoretically desirable but also
experimentally feasible. That is by no means an \emph{a priori} certainty as
the expected dense interspecies Feshbach resonance structure might easily have
interdicted such a state by strong inelastic collisional losses at the low
magnetic fields that are typically used during standard evaporation stages.
Along the way, our experiment once again underlines the quite amazing
versatility of the heavy Yb not only from a physics standpoint but also in its
capability to efficiently cool a variety of other species
sympathetically~\cite{hansen_quantum_2011, schafer_feshbach_2022}.

The present work focused on the most commonly used bosonic isotopes
of Er and Yb. An extension to different isotope combinations with
interesting applications to fundamental research~\cite{kosicki_quantum_2020}
appears straightforward and our experiment is already set up to also work
with various bosonic and fermionic isotopes.

It seems obvious that from these encouraging results it is only a few
minor steps towards a full investigation of the \bEr-\bYb\ interspecies
Feshbach resonance spectrum for at least this isotope combination. In the case
of our experiment, in particular, we will, however, need to first make some
improvements to (i) increase the atoms numbers in the final ultracold samples
and (ii) improve the magnetic field stability of our
setup~\cite{schafer_feshbach_2022}. The latter will be necessary to reliably
identify also the narrow Feshbach resonances in order to obtain complete
Feshbach spectra that are necessary for the envisioned statistical analyses.
These instrumental challenges are not insurmountable, and we are therefore
confident that our results will further enrich the family of experiment-proven
ultracold atomic mixtures to address some of the quantum questions of our
times.

\section*{Acknowledgments}
This work was supported by the Grant-in-Aid for Scientific Research of JSPS
Grants No.\ JP17H06138, No.\ 18H05405, and No.\ 18H05228, JST CREST Grant No.\
JPMJCR1673 and the Impulsing Paradigm Change through Disruptive Technologies
(ImPACT) program by the Cabinet Office, Government of Japan, and MEXT Quantum
Leap Flagship Program (MEXT Q-LEAP) Grant No.\ JPMXS0118069021 and Moonshot
Program Grant No.\ JPMJMS2269.

\end{document}